\documentclass{pasj00}
\usepackage{graphicx,color}

\begin{document}
\SetRunningHead{Tsunoda et al. 2006}{Accretion disk in M81 X-9}
\Received{2006/02/24}
\Accepted{2006/10/12}

\title{Detailed spectral study of an ultra-luminous compact X-ray source M81 X-9 in the disk dominated state}
\author{Naoko \textsc{Tsunoda},\altaffilmark{1,2}
	Aya \textsc{Kubota},\altaffilmark{1}
	Masaaki \textsc{Namiki},\altaffilmark{3}
	Masahiko \textsc{Sugiho},\altaffilmark{4}\\
	Kiyoshi \textsc{Kawabata},\altaffilmark{2} 
	and Kazuo \textsc{Makishima}\altaffilmark{1,5}}%
\altaffiltext{1}{Institute of Physical and Chemical Research (RIKEN), 2-1 Hirosawa, Wako, Saitama 351-0198}
\altaffiltext{2}{Department of Physics, Tokyo University of Science, 
                 1--3 Kagurazaka, Shinjuku--ku, Tokyo 162--8601}
\email{tsunoda@crab.riken.jp}
\altaffiltext{3}{Department of Earth and Space Science, Osaka University,\\
		1--1 Machikaneyama--cho, Toyonaka, Osaka 560--0043}
\altaffiltext{4}{NEC TOSHIBA Space Systems, Ltd.
1-10, Nissin-cho, Fuchu, Tokyo, 183-8551, Japan}
\altaffiltext{5}{Department of Physics, University of Tokyo,
		7--3--1 Hongo, Bunkyo--ku, Tokyo 113--0033}

%

\KeyWords{accretion, accretion disks---black hole physics---X-rays: individual (M81 X-9, Holmberg IX X-1)---X-rays: stars} 

\maketitle

\begin{abstract}

We report on the results of detailed spectral studies of
the ultra-luminous X-ray source (ULX),
M81 X-9 (Holmberg IX X-1), made with
XMM-Newton on 2001 April 22 and with ASCA on 1999 April 6.
On both occasions, the source showed an unabsorbed 0.5--10 keV luminosity of 
$\sim2\times10^{40}~{\rm erg~s^{-1}}$(assuming a distance of 3.4~Mpc) and 
a soft spectrum apparently represented by a multi-color disk model 
with an innermost disk temperature of 1.3--1.5~keV. 
Adding a power-law model further improved the fit.
However, as previously reported, the high luminosity cannot be reconciled with the high disk temperature within a framework of the standard accretion disk radiating at a sub-Eddington luminosity.
Therefore, we modified the multi-color disk model, and allowed the local disk temperature to scale as $\propto~r^{-p}$ on the distance $r$ from the black hole, with $p$ being a free parameter.
We then found that the XMM-Newton and the ASCA spectra can be both reproduced successfully with $p \sim 0.6$ and the innermost disk temperature of 1.4--1.8 keV.
These flatter temperature profiles 
suggest deviation from the standard Shakura-Sunyaev disk, and
are consistent with predictions of a slim disk model.

\end{abstract}

\section{Introduction}

Nearby spiral galaxies often host ultra-luminous compact X-ray sources 
	(hereafter ULXs; Makishima et al. 2000), 
	with X-ray luminosities reaching 
	$3\times 10^{39}$--$10^{40}~{\rm erg~s^{-1}}$.
	Although they require black hole masses 
	of 20--100~$M_\odot$ or more  by 
	considering the Eddington limit, $L_{\rm E}$, 
	it is not obvious whether such massive stellar black holes can be 
	created as an end point of evolution of massive stars.
Thus, there are still extensive arguments on the nature of the ULXs.
Thanks to ASCA, a great leap has been achieved
	in our understanding of 1--10 keV spectra of a dozen ULXs.   
As reported by many authors (\cite{okada98,Makishima2000}, and references therein), 
	spectra of a majority of the ASCA sample have been reproduced 
	by so-called multi-color disk (MCD) model ({\sc diskbb} 
	in {\sc xspec}\footnote{http://heasarc.gsfc.nasa.gov/docs/xanadu/}; 
	\cite{Mitsuda1984}; \cite{Makishima1986}), 
	while others are described with a power-law model.
	The MCD model approximates emission from a geometrically thin, 
	radiation dominated standard accretion disk (Shakura \& Sunyaev 1973),
	and successfully describes the dominant soft spectral component 
	of Galactic/Magellanic black hole binaries 
	(hereafter BHB) in the high/soft state. 
 In addition, spectral transitions between the disk dominant type and 
	the power-law dominant type states have been observed in several ULXs 
	(e.g., \cite{kubota02}; \cite{La Parola2001}).
These spectral characteristics of ULXs have thereby reinforced 
	their candidacy as accreting black holes.

In spite of the successful application of the MCD model 
	to the ULX spectra, there still remain three self inconsistencies.
One is known as ``too high $T_{\rm in}$" problem that 
	the values of the innermost disk temperature, 
	$T_{\rm in}$, obtained from the disk-dominant ULXs 
	(typically 1--2 keV; \cite{Makishima2000}) 
	are significantly higher than those of BHBs (0.5--1.2 keV), 
	even though 
	higher-mass black holes should have lower values of $T_{\rm in}$
	as $T_{\rm in}\propto M^{-1/4}$ 
	under the standard-disk picture (e.g., \cite{Makishima2000}).
The second one is known as ``variable $r_{\rm in}$" problem: 
        the inner disk radius, $r_{\rm in}$, of ULXs are reported
	to vary as $r_{\rm in}\propto T_{\rm in}^{-1}$ 
	(\cite{Mizuno2001}), 
	in contradiction to the case of high-state BHBs in which $r_{\rm in}$
	is generally constant at a value consistent with 
	the radius of last stable orbit around the central black hole
	(e.g., \cite{Makishima1986}; \cite{tanaka95};
	\cite{Ebisawa1993}; \yearcite{Ebisawa1994}).
The last problem is that the ULXs luminosity at which 
	the spectral transition occurs, $\sim 10^{39}~{\rm erg~s^{-1}}$, 
	is too high to be considered as 
	the classical high-low state transitions of BHBs 
	which generally occurs at several percent of $L_{\rm E}$

The ``too high $T_{\rm in}$" problem might be solved within the framework 
	of standard accretion disks, by invoking extreme assumptions.
For example, a rapid black hole rotation could increase $T_{\rm in}$ 
	(\cite{Makishima2000}) but the effect may not be sufficient 
	to explain the ULX spectra (\cite{Ebisawa2003}).
Alternatively, the values of $T_{\rm in}$ could increase 
	if the color-to-effective correction factor $\kappa$ took 
	a very high values as 3--5, for some unknown reasons,
	but this would contradict the canonical value of $\kappa = $1.5--2.0
	suggested theoretically (\cite{Shimura1995}; \cite{Davis2005}) and 
	observationally (\cite{Makishima2000}).
Furthermore neither of these ``solutions" may solve the second and third problems.

Considering these difficulties,
        accretion flows in ULXs may be significantly deviated from 
	the description of standard accretion disks, 
	even though the emergent spectra look rather alike.
Actually, when the mass accretion rate, $\dot{M}$, increases and 
	the luminosity becomes closer to $L_{\rm E}$, 
	the disk structure is expected to change from the standard disk to 
	a slim disk (e.g., \cite{Abramowicz1988,Watarai2000}).
The slim disk is predicted to have a significantly higher inner temperature
	than the standard accretion disk, and its apparent 
	$r_{\rm in}$ (when fitted with the MCD model) is expected to 
	vary roughly as $\propto~T_{\rm in}^{-1}$  (\cite{Watarai2000})
	just as is observed.
	These explain away the first and second problems, respectively (e.g., \cite{Mizuno2001,Watarai2001}).
Further, invoking the slim disk implies that the radiation luminosity 
	is close to $L_{\rm E}$, thus solving the third problem.
Therefore, the slim disk scenario can potentially solve the three problems
	associated with the interpretation of disk dominated ULXs.

While the radiation-dominated standard disk is characterized by a 
	local temperature varying as $T(r)\propto r^{-3/4}$ 
	({\it r} being the distance from the central black hole), 
	the temperature profile of the slim disk is predicted 
	to be flatter than this as $T(r) \propto r^{-0.5}$.
The different temperature profiles are expected, in turn, to give 
        subtle differences to the overall disk spectra.
This effect may be described by
	so-called $p$-free disk model (a local model for 
	{\sc xspec}\footnote{http://heasarc.gsfc.nasa.gov/docs/xanadu/xspec/models/diskpbb.html}; \cite{Mineshige1994}; \cite{Hirano1995}; 
	\cite{Kubota2004}),
	it assumes that the spectrum  is composed of multi-temperature 
	blackbody emission, where the local temperature $T(r)$  
	at a radius {\it r} is given by $T(r)=T_{\rm in}(r/r_{\rm in})^{-p}$, 
	with {\it p} being a positive free parameter.
Using the ASCA and RXTE data of Galactic/Magellanic BHBs, 
	it has been 
        confirmed that {\it p} takes a value consistent with 
	the standard value, $3/4$, when the objects are considered
	to reside in the standard-disk state with constant values 
	of $r_{\rm in}$ (\cite{Kubota2005}), 
	and that {\it p} becomes smaller than 3/4
	in 
	a supposedly slim disk state (\cite{Kubota2004};
	\cite{Abe2005}).

In the present paper, we apply the {\it p}-free disk model analysis to 
	X-ray spectra of M81 X-9 in the disk dominant state, obtained 
	with XMM-Newton on 2000 April 22 and with ASCA 
	on 1999 April 6, 
	in search for the expected spectral deviation from the standard disk.
At a distance of 3.4~Mpc (\cite{Georgiev1991}),
	this source is an extensively studied ULX
	associated with the dwarf galaxy Holmberg IX,	
	which is a close companion to M81.
Through previous snapshot observations with ASCA in 1993--1999,  
	its unabsorbed 0.5--10~keV luminosity was found in  
	(6--20)$\times 10^{39}~{\rm erg~s^{-1}} $ assuming 
	an isotropic emission.
Until 1998, the source kept a relatively low luminosity of 
	$ \sim 6 \times 10^{39}~{\rm erg~s^{-1}} $,
	with a single power-law spectrum of photon index  
	$\Gamma \sim 1.4 $ \citep{Ezoe2001}.
In the 1999 observation, the source was reported to have made  
	a spectral transition to a disk-dominant state 
	with $T_{\rm in} = 1.24$ keV, together with an increased luminosity of 
	$ \sim 1.8 \times 10^{40}~{\rm erg~s^{-1}}$ (\cite{La Parola2001}).
Good XMM-Newton observations of M81 X-9 were performed 
        three times by 2002.
In 2001, its luminosity and spectral shape were almost 
	the same as those obtained in the 1999 
	ASCA observation (Sugiho~2003).
In the 2002 observations, the source became slightly fainter 
	with an unabsorbed  0.3--10~keV luminosity of
	$ L = 1.1$--$1.3 \times 10^{40}~{\rm erg~s^{-1}} $, 
	and its spectrum returned to the power-law dominant state 
	with a hint of cool MCD component (\cite{Miller2004}).

Out of these multiple datasets, here we select two particular ones
        for our {\it p}-free study, namely
	the 1999 ASCA data and the 2001 XMM-Newton data,
	because the object was clearly in the disk dominant state 
	on these occasions.
These data sets are expected 
to give a significantly smaller values of $p$  than that 
of standard disk, 3/4, if the slim disk is realized.
These are the same data as analyzed already by La Parola et al. (2001) 
	and Sugiho (2003), respectively.

\section{Observations and Data Reduction}

The present XMM-Newton observation of M81 X-9 was performed for 
        127.9 ks on 2001 April 22  (observation ID of 0111800101), with the three detectors of
	the European Photon Imaging Camera (EPIC).
The two MOS cameras were operated in full frame mode, 
        and the PN camera with the medium filter was operated in small
	window mode.
We retrieved the data from the XMM-Newton public archive\footnote{http://xmm.vilspa.esa.es/}, 
and
used only MOS-1 data for the spectral study, since the source fell in the chip boundary of MOS-2 and outside the field of view (FOV) of the PN camera.
The event file was reprocessed and filtered with the XMM-Newton Science Analysis Software (SAS v6.0.0). 
Utilizing light curves of source-free regions from the MOS-1 data, we rejected intervals with high background levels, and obtained a net exposure of 79.7 ks for the MOS-1 camera. 

\begin{figure}
  \begin{center}
    \FigureFile(80mm,80mm){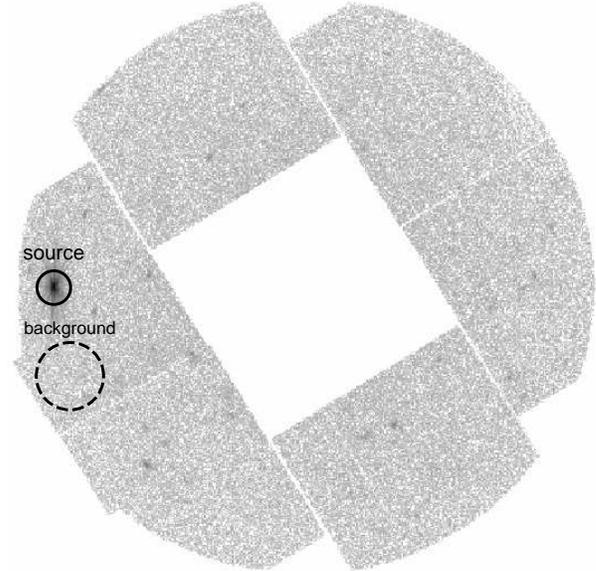}
  \end{center}
  \caption{A 0.5--10 keV unsmoothed MOS-1 image of the M81 X-9 field obtained on 2001 April 22. The source region and the background region are shown with a solid circle and a dotted circle, respectively. The background is included.}
  \label{fig:X-9_image}
\end{figure}

\begin{figure}
  \begin{center}
    \FigureFile(80mm,80mm){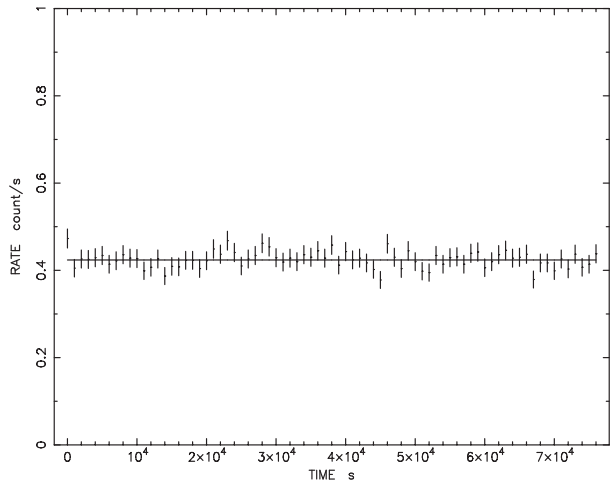}
  \end{center}
  \caption{The 0.5--10 keV MOS-1 light curve of M81 X-9, plotted with a time bin of 1~ks.  
A solid line indicates the best fit constant.}
  \label{fig:lc}
\end{figure}

Figure \ref{fig:X-9_image} shows a MOS-1 image in the range of 0.5--10 keV.
The source is thus located at \timeform{12'.5} off the center of the MOS-1 FOV.
 Using a SAS task {\sc xmmselect} (v6.0.1), 
        the MOS-1 events with patterns 0 to 12 were extracted 
	from a circular region of \timeform{50"} radius around 
	the position of M81 X-9,  
	($\alpha,\delta)=(\timeform{09h57m53s},+\timeform{69D03'47"}$).
Background events were accumulated over a source-free region 
        with a $\timeform{1'.7}$ radius, at a position
	($\alpha,\delta)=(\timeform{09h57m43s},+\timeform{68D59'24"}$).
The on-source and the background regions are indicated 
        in figure \ref{fig:X-9_image}.
Figure \ref{fig:lc} shows the 0.5--10 keV light curve of M81 X-9. 
There is no significant intensity variations 
beyond the Poisson noise, with a typical upper limit of 10\%.
The average count rate, $0.424 \pm 0.004~\rm{cts~s^{-1}}$,
        roughly corresponds to a 0.5--10 keV flux of 
	$1.4 \times 10 ^{-11}~\rm{erg~s^{-1}~cm^{-2}}$.

Since M81 X-9 is located close to the edge of MOS-1 FOV, 
        its spectral analysis might be subject to calibration 
	uncertainties.
Therefore, we analyzed three data sets of a calibration target, 
        the supernova remnant G21.5-09 which shows a power-law spectrum, 
	one acquired near the FOV center and the others 
	$\timeform{11'.12}$ and $\timeform{13'.02}$ off.
As detailed in Appendix, we have confirmed that the spectra of such 
        off-axis sources can be studied with a high reliability.
        Though the off-axis data may yield slightly harder spectrum than the on-axis ones, 
        the difference in the photon index is less than $\sim0.1$ (see  figure6 in Appendix).

To compare with the XMM-Newton result on the disk dominant state, 
        we also reanalyzed the ASCA 
SIS (Solid state Imaging Spectrometer) data of M81 X-9 
obtained on 1999 April 6. 
We used the screened data set provided by the DARTS science  
archive\footnote{http://www.darts.isas.jaxa.jp/index.html} with  
a net exposure of 33~ks.
The source spectra were accumulated over a circular region centered on the source, with a radius of $\timeform{2'.9}$, and the background spectra were created using blank sky data.
The criteria of the data screening are as follows:
a) the object be at least 10$^\circ$ above the night Earth's limb or
 at least 20$^\circ$ above the bright Earth's limb;
c) the cutoff rigidity (COR) of cosmic rays be greater than $6~{\rm GeV~c^{-1}}$;
and d) the time after day night transition be greater than 100 s.
The 0.5--10 keV count rate of M81 X-9 was  0.326$\pm$0.003 and
0.272$\pm$0.003 $\rm {cts~s^{-1}}$ with SIS0 and SIS1,
respectively.
The implied 0.5--10 keV flux, approximately 
$ 1.3 \times 10^{-11}~{\rm erg~s^{-1}cm^{-2}} $,
is close ($\sim 93$ \%) to that of the 2001 XMM-Newton data.

\section{Data Analysis and Results}

\subsection{Analysis of the XMM-Newton data with canonical models}

A left panel of  figure \ref{fig:spec} shows a MOS-1 spectrum of M81 X-9 in the range of 0.5--10 keV.
Using a SAS task {\sc epatplot}, the MOS-1 data were confirmed to be
free from photon pile-up effects.
Following the standard modeling of ULXs (e.g., Makishima et al. 2000),
we first examined the spectrum with a single power-law model and the MCD
model, and obtained the fitting results as shown in Table~\ref{ta:fitting}.
Here, the response matrices were created by {\sc rmfgen} and {\sc
arfgen} (v6.0.1), and the spectral fitting was performed with 
the software package {\sc xspec} 11.2.0. 
The model was subjected to photoelectric absorption by a hydrogen
column which is left free to vary.
The single power-law fit was far from acceptable, with
$\chi^2/(\rm{d.o.f}) = 574.1/335 $.
The single MCD fit was much more successful, and yielded the parameters
which are very similar to the ASCA results in 1999 reported by \citet{La Parola2001}. 
However, the fit still remained unacceptable with $\chi^2/(\rm{d.o.f}) = 427.4/335 $.

We hence fit the data with an MCD plus power-law model, as is done for galactic BHBs in the high state. 
This two-component model has successfully reproduced the observed spectrum with $\chi^2/(\rm{d.o.f}) = 359.4/333 $.
As shown in Table~\ref{ta:fitting}, the best-fit model implies 
an absorbed 0.5--10 keV flux of $1.4 \times 10^{-11} ~\rm{erg~s^{-1}cm^{-2}}$, in which the MCD component accounts for $\sim 60$\%.
The unabsorbed luminosity in the same band is obtained as $2.3 \times 10^{40} ~\rm{erg~s^{-1}}$, assuming an isotropic emission.
The MCD parameters, $T_{\rm in}$ and $r_{\rm in}$, 
have been measured to be $1.45$ keV  and $(160\pm12) \cdot \alpha$~km
with $\alpha=\sqrt{\cos 60^\circ/\cos i}$ ({\it i} being the inclination angle of the disk), respectively.
The true inner radius of the disk is calculated as 
$R_{\rm in}=(190\pm15) \cdot \alpha$~km,
by using a relation of $R_{\rm in}=\xi\kappa^2 r_{\rm in}$, when
$\xi=0.41$ is
a correction factor for the inner boundary condition (\cite{Kubota1998})
and $\kappa=1.7$ is a spectral hardening factor
(\cite{Shimura1995}). 
The disk bolometric luminosity is estimated as 
$L_{\rm bol}=1.5\times10^{40}\cdot \alpha^2~{\rm erg~s^{-1}}$
through a relation of $L_{\rm bol}=4\pi \sigma r_{\rm in}^2 T_{\rm in}^4$ (\cite{Makishima1986}),
with $\sigma$ the Stefan-Boltzman constant. 
When the 0.5--10~keV unabsorbed power-law luminosity,
$8.7\times10^{39}~{\rm erg~s^{-1}}$, is added to the disk bolometric
luminosity, the total luminosity is roughly estimated as $2.4\times10^{40}~{\rm erg~s^{-1}}$ for $\alpha=1$ (i.e., $i=60^\circ$).

The observed X-ray luminosity requires a black hole mass of $>150~M_\odot$ 
if the Eddington limit is satisfied.
In contrast, 
the observed disk temperature of 1.45~keV
is too high for the emission to be coming from a standard disk around such a massive black hole.
In other words, 
the disk inner radius of 190~km is too small to be identified
with the last stable orbit of a non-spinning black hole with $>150~M_\odot$.
If the obtained value of $R_{\rm in}$ were required to coincide with $6R_{\rm g}$, then the Schwarzshild radius, $2R_{\rm g}$, of this black
hole would be only $\sim64$~km, which is equivalent to a black hole mass of
$\sim20~M_\odot$. 
Thus, the 2001 XMM-Newton observation of M81 X-9 clearly
reproduces the ``too high $T_{\rm in}$'' problem described in \S~1.

\subsection{Analysis of the ASCA data in 1999}

The flux and the MCD parameters obtained from the 2001 XMM-Newton
data (\S~3.1) are very similar to those reported by \citet{La Parola2001}, who analyzed the ASCA SIS data obtained on 1999 April 6 
with the single power-law model and
the single MCD model as introduced in \S ~3.1.
In order to compare the ASCA and XMM-Newton spectra in a
unified manner, we reanalyzed the same 0.5--10 keV ASCA SIS data
with the same two-component model.

A right panel of  figure \ref{fig:spec} shows the ASCA SIS (SIS0 and SIS1) 
spectra with the best fit models. 
The results of this analysis are also shown in Table~\ref{ta:fitting}.
The obtained best-fit parameters of the single power-law
 fit and the single MCD fit are consistent with those reported by \citet{La Parola2001}.
Although the single MCD model gave an acceptable fit with $\chi^2/({\rm
d.o.f}) =318/301$, addition of a power-law component improved the fit
significantly in terms of an $F$-test, with $F(2, 301)=20.8$ which corresponds
to 99.99~\% significance.
This two-component fit implies a 0.5--10 keV flux of $1.3 \times10^{-11} \rm{erg~s^{-1}cm^{-2}}$ 
with the MCD component contributing $\sim 70$\%, 
and an unabsorbed 0.5--10 keV luminosity of $2.4 \times 10^{40} \rm{erg~s^{-1}}$.
The values of $T_{\rm in}$ and $r_{\rm in}$ have been estimated as 
$1.28^{+0.12}_{-0.14}$~keV and 
$(220\pm50)\cdot \alpha$~km, 
respectively 
(Table \ref{ta:fitting}). 
The spectral shape and the luminosity are thus almost the same, within
errors, between
the two observations with the different instruments.

\subsection{Temperature profile of the optically thick disk}\label{p-free}

\begin{figure*}
  \begin{center}
	\begin{minipage}{80mm}
    \FigureFile(80mm,80mm){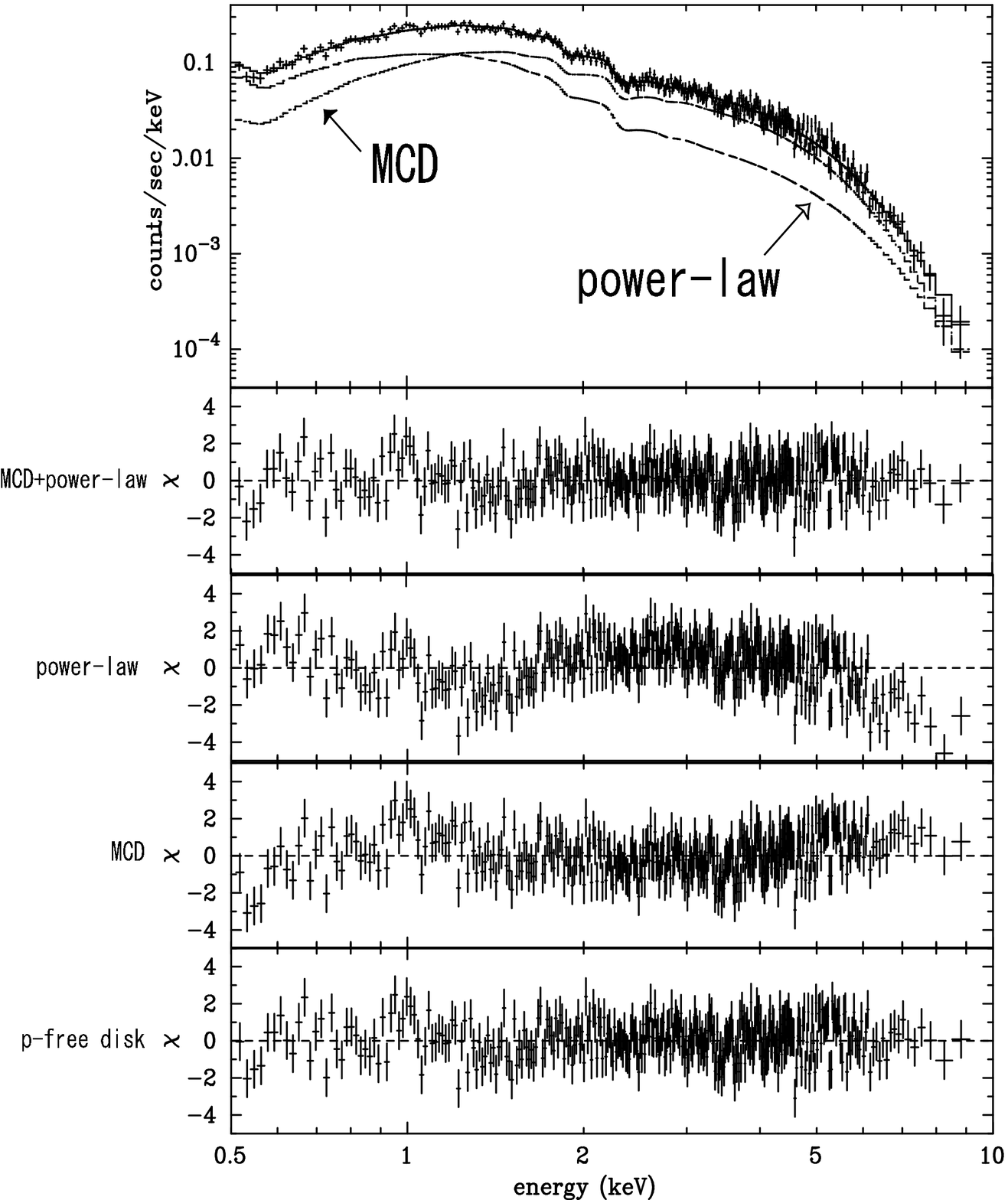}
	\end{minipage}
	\begin{minipage}{80mm}
    \FigureFile(80mm,80mm){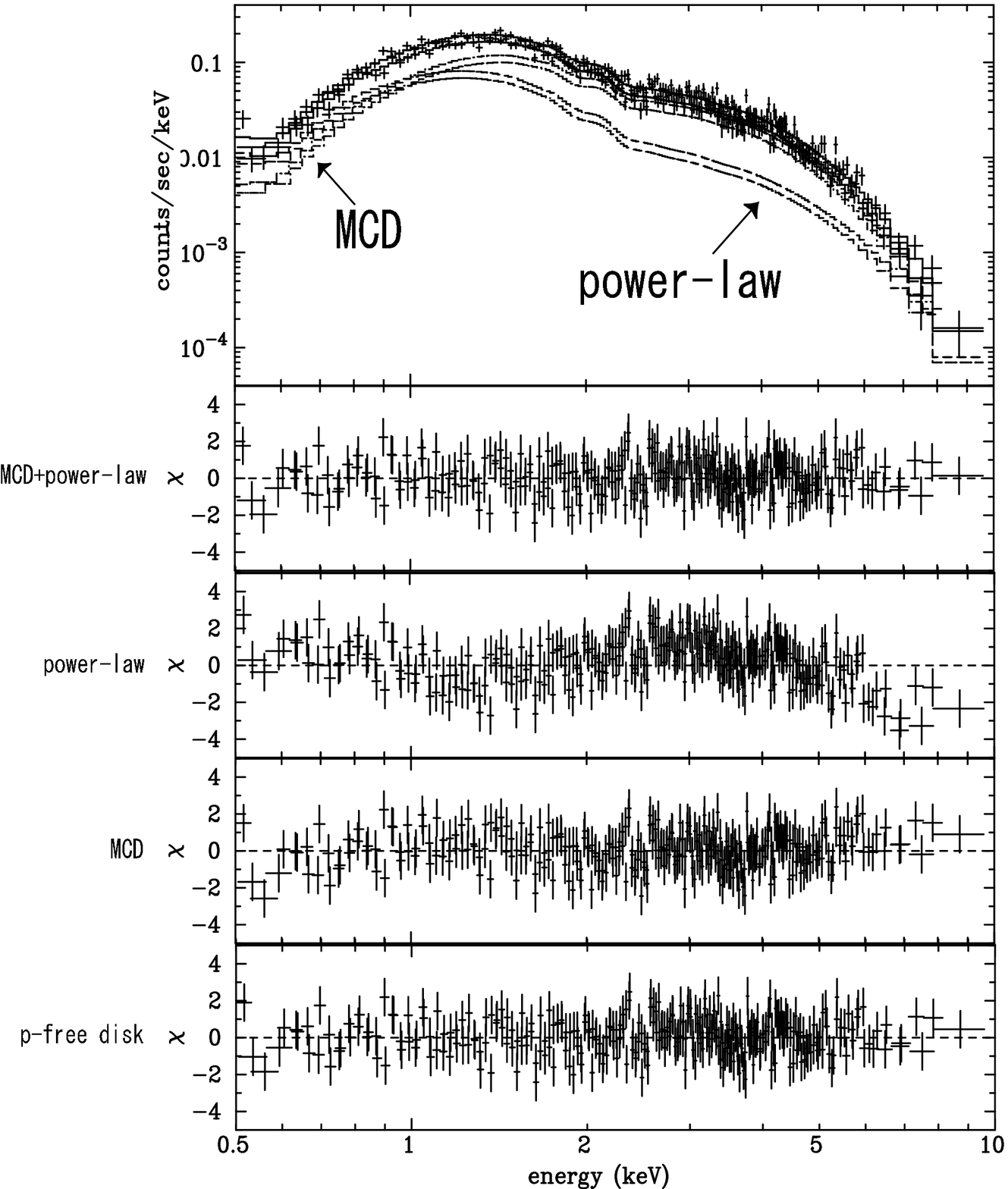}
	\end{minipage}
  \end{center}
  \caption{The 0.5--10~keV spectra obtained with the XMM-Newton MOS-1 on 2001 April 22 (left), and with the ASCA SIS (SIS0 and SIS1) on 1999 April 6 (right). The best-fit MCD plus power-law models are superposed on the data. Bottom four panels show residuals between the data and the attempted models; from top to bottom, the MCD plus power-law, single power-law, single MCD, and the $p$-free disk models.}
  \label{fig:spec}
\end{figure*}

In addition to the ``too high $T_{\rm in}$'' problem, 
we encountered another problem in the spectral fit of M81 X-9. 
Although the two-component (MCD plus power-law) model apparently reproduces the spectra from the ASCA 
and the XMM-Newton observations,
the power-law component exceeds the MCD component at
energies below $\sim$ 1 keV, as shown
 in  figure \ref{fig:spec}.
Such a fit would be rather unphysical, 
since the power-law tail is generally considered as a consequence of
inverse-Compton scattering of the MCD photons by high-energy electrons
 created around the accretion disk.
Therefore, the hard-tail flux should not dominate that of the seed MCD,
at least in the energy band below $2.8~kT_{\rm in}$ which is the peak
energy of a blackbody emission of temperature $T_{\rm in}$.
The apparent low-energy dominance of the power-law component suggests a
subtle deviation of the accretion disk configuration from the standard
one, in such a way that the softest end of the radially-integrated
spectrum is more enhanced than the MCD model.

As described in \S1, 
such a spectral change may arise if, e.g., the radiative efficiency of the 
inner disk region becomes reduced and the radial temperature gradient 
flattens; 
such effects can be quantified using the $p$-free disk model (\cite{Kubota2004}).
We hence fitted the MOS-1 and SIS data with the {\it p}-free disk model,
and obtained the results 
presented in Table \ref{ta:fitting}
and  figure \ref{fig:spec}.
In both the XMM-Newton and ASCA spectra, the $p$-free disk model
is similarly, or even more, successful than the MCD+power-law fit.
The values of {\it p} have been found as 
$0.60\pm0.02$ and $0.61^{+0.04}_{-0.03}$ with
the {\it Newton} and ASCA data, respectively, in a very good
mutual agreement.
They are significantly smaller 
than the standard value of $p=3/4$.
The absorption column becomes somewhat higher than that implied 
by the single MCD fit.
To reinforce the result, we simulated the spectrum by considering  
the possible systematic difference in the spectral index of 
$\Delta\Gamma\approx 0.1$ for the Newton off-axis sources (see  figure~6 in Appendix). 
This difference was found to make the best fit value of $p$ slightly smaller as $p=0.58\pm0.02$, but it is within the 
statistic uncertainty.

\begin{table*}
  \caption{Spectral fitting parameters and derived quantities of M81 X-9.\footnotemark[$*$].}
  \label{ta:fitting}
  \begin{center}
    \begin{tabular}{lcccccccc}
\hline
  Model &  ${N_{\rm{H}}}$\footnotemark[$\dagger$] 
        & $\rm{\Gamma}$ or {\it p}\footnotemark[$\ddagger$] 
        & $T_{\rm{in}}$~(keV) & $r_{\rm in}{~(\rm km)}$\footnotemark[$\S$] 
	& $M_{\rm MCD}~(M_{\odot})$\footnotemark[$\S\S$]  & $F_{\rm{X}}$\footnotemark[$\|$]
	& $L_{\rm bol}$\footnotemark[$\#$] 
	& $\rm{{\chi}^2 / d.o.f.}$ \\
\hline\hline
  \multicolumn{9}{c}{2001 April 22~~~~(XMM-Newton)}\\ 
\hline\hline
  power-law & 2.88$_{-0.12}^{+0.11}$ & 1.96$_{-}^{+}$0.03  &  
	& & & 14.9 & & 574.1/335 \\
  MCD & 0.75$_{-0.07}^{+0.08}$ & & 1.35$_{-0.03}^{+0.02}$ 
	& 220 $\pm$ 4  
	& 30 & 13.0 & 21.0 & 427.4/335\\ 
  MCD & 2.17$_{-0.42}^{+0.63}$ & & 1.45$_{-0.14}^{+0.10}$ 
	&160 $\pm$ 12  
	& 22 & 8.4 & 14.6 & 359.4/333 \\
  ~~+power-law & & 2.28$_{-0.35}^{+0.58}$ & & & &  5.1 & & \\ 
  {\it p}-free disk & 1.70$_{-0.20}^{+0.19}$ & 0.60$_{-}^{+}$0.02  
              & 1.72$_{-0.10}^{+0.11}$ & 100 $\pm$ 20&---& 13.5 
		& 28.4\footnotemark[$\dagger\dagger$] & 357.3/334  \\ 
\hline\hline
\multicolumn{9}{c}{1999 April 06~~~~(ASCA)}\\ 
\hline\hline
  power-law & 4.68$_{-0.23}^{+0.24}$ & 2.16$_{-}^{+}$0.04 &  
            & & & 13.9 & & 453.3/301\\
  MCD & 1.71$_{-}^{+}$0.14 &  & 1.25$_{-}^{+}$0.03 
	& 260 $\pm$ 13  
	& 35 & 12.4 & 21.8 & 318.5/301\\
  MCD & 3.18$_{-0.94}^{+2.76}$ &  & 1.28$_{-0.14}^{+0.12}$ 
	& 220 $\pm$ 50 
	& 29 & 8.5 & 16.1 & 297.8/299\\
  ~~+power-law &  & 2.35$_{-0.65}^{+1.95}$ &  & & & 4.3 & &  \\
  {\it p}-free disk & 2.68$_{-0.38}^{+0.37}$ & 0.61$_{-0.03}^{+0.04}$ 
              & 1.47$_{-0.10}^{+0.12}$ & 140 $\pm$ 30& ---& 12.6 
		& 28.3\footnotemark[$\dagger\dagger$] & 298.3/300 \\ 
\hline
   \multicolumn{9}{l}{\footnotesize
	\par\noindent
	\footnotemark[$*$] All quoted uncertainties are at the 90 \% confidence limit.}\\
   \multicolumn{9}{l}{\footnotesize 
	\par\noindent
	\footnotemark[$\dagger$] Hydrogen column density for the photoelectric 
             absorption, in units of $\rm{10^{21}~cm^{-2}}$.}\\
   \multicolumn{9}{l}{\footnotesize
	\par\noindent
	\footnotemark[$\ddagger$] The photon index for the power-law
     model, and the value of {\it p} for the {\it p}-free disk 
             model.}\\
   \multicolumn{9}{l}{\footnotesize
	\par\noindent
	\footnotemark[$\S$] An apparent innermost disk radius, assuming a distance of 
             $D=3.4$ Mpc and a disk inclination of $i=60^{\circ}$. }\\ 
      \multicolumn{9}{l}{\footnotesize
	\par\noindent          
            \footnotemark[$\S\S$] Black hole mass estimated by assuming that  
            the true inner radius $R_{\rm in}$ coincides with $6R_{\rm g}$. }\\
          \multicolumn{9}{l}{\footnotesize
	\par\noindent                  
            Here the value of $R_{\rm in}$ is estimated by $r_{\rm in}$ with correction factors of $\xi=0.41$ and $\kappa=1.7$.}\\ 
   \multicolumn{9}{l}{\footnotesize
	\par\noindent
	\footnotemark[$\|$] The absorption-uncorrected flux,  
           in units of $\rm{10^{-12}~erg~s^{-1} cm^{-2}}$. }\\
   \multicolumn{9}{l}{\footnotesize
	\par\noindent
	\footnotemark[$\#$] The bolometric luminosity, in units of 
             $10^{39} \rm{erg~s^{-1}}$.}\\
   \multicolumn{9}{l}{\footnotesize
	\par\noindent
	\footnotemark[$\dagger\dagger$] The 0.01-100 keV luminosity estimated 
            from the best-fit model assuming isotropic emission. }\\
    \end{tabular}
  \end{center}
\end{table*}


%
%


\newpage

\section{ Discussion}

Using the ASCA data acquired on 1999 April 6
and the XMM-Newton data on 2001 April 22,
we studied the X-ray spectra of M81 X-9.
On both occasions,
the source exhibited typical disk dominated spectra,
with very similar sets of spectral parameters
including the bolometric luminosity of
$\sim 2 \times 10^{40}$ erg s$^{-1}$.
Combined with the power-law type spectra
observed from this object at somewhat (a factor of $\sim 2$)
lower luminosities
(\cite{Ezoe2001}; \cite{Miller2004}),
our results reinforce the reports by
\citet{La Parola2001} and \citet{Sugiho2003}
that M81 X-9 makes occasional transitions between
the disk dominant and power-law dominant states.

In  figure 4,
we plot  the present two measurements
on the plane of $L_{\rm disk}$ vs. $T_{\rm in}$ after \citet{Makishima2000}.
For comparison, we also show results
on several high-$T_{\rm in}$ ULXs,
referring to Mizuno et al.(2001).
Thus, the ULX data points, including those of M81 X-9,
mostly fall on the apparent ``super-Eddington'' region,
where $L_{\rm bol}$ significantly exceeds
$L_{\rm E}$  which is calculated assuming
standard accretion disks with $R_{\rm in}=6 R_{\rm g}$.
This is just another representation of
the ``too high $T_{\rm in}$'' problem.
The figure also reveals the ``variable $r_{\rm in}$'' problem,
because the ULX variations do not trace the dotted lines
representing $L_{\rm bol} \propto T_{\rm in}^4$,
but instead,
behave approximately as  $L_{\rm bol} \propto T_{\rm in}^2$
(or equivalently $r_{\rm in} \propto T_{\rm in}^{-1}$).
Thus, we consider
that M81 X-9 is subject to the same problems as described in \S~1,
although its variation is not very prominent on the diagram.

In order to solve these problems with the standard-disk
interpretation of  disk dominant spectra of luminous ULXs,
we have adopted the slim-disk scenario as a working hypothesis,
and analyzed the two data sets of M81 X-9 using the $p$-free disk model.
As a result,
the XMM-Newton and ASCA spectra have both  been described
by this model more successfully than by the traditional MCD model.
Furthermore,  the derived temperature profiles,
$p=0.60 \pm 0.02$ with XMM-Newton
and $p=0.61^{+0.04}_{-0.03}$ with ASCA,
agree with each other,
and are significantly smaller than the canonical
value of $p=3/4$ to be found with standard disks.
In view of the theoretical prediction
by \citet{Watarai2000} described in \S~1,
these  results strongly suggest
the presence of a slim disk in M81 X-9 on these occasions.
In particular, the good agreement between
the two different instruments significantly reduces the possibility
that the deviation of $p$ from 3/4 is an instrumental artifact.

In  figure 4, results on the Galactic BHB, XTE J$1550-564$,
are also presented for comparison.
At luminosities below $\sim  0.4 L_{\rm E}$,
the data points of this BHB thus trace tightly a constant-mass
(or $L_{\rm bol} \propto T_{\rm in}^4$) locus.
At  higher luminosities, however,
the source starts deviating from this relation \citep{Kubota2004},
exhibiting the ``variable $r_{\rm in}$'' problem just like the ULXs.
The same behavior has been observed  from other BHBs as well,
including 4U~$1630-47$ \citep{Abe2005}
and GRO~J$1655-40$ \citep{Kubota2001}.
\citet{Kubota2004} named this characteristic
spectral state of BHBs {\it apparently standard regime},
because the spectra in this state resemble those from standard disks,
except for significantly higher values of $T_{\rm in}$.
They showed that the  RXTE spectra of XTE J$1550-564$  in this state
can be described by the $p$-free model with  $p < 0.75$.

Based on these considerations, we may presume
that  high-$T_{\rm in}$ ULXs
and Galactic BHBs in the {\it apparently standard regime}
have essentially the same spectral characteristics,
which in turn are explained in terms of the formation of a slim disk;
the only difference is
that  the ULXs are by an order of magnitude more luminous.
Then, the first problem (too high $T_{\rm in}$) can be solved at 
least qualitatively,
because a slim disk appears significantly hotter than a standard disk.
This interpretation also solves
the second problem  (variable $r_{\rm in}$),
because such a behavior is predicted by the slim disk theory
(Watarai et al. 2000).


Then, how about the third problem, 
too high transition luminosity between the disk dominant and the power-law dominant 
states?
Luminosity differences associated with ULX transitions (a factor of $\sim 2$ in the particular case of M81 X-9) are generally too small  to  identify the transitions with those between the low and high states of BHBs.
In addition to the classical high-low transition, 
BHBs make  transitions between the {\it standard regime}
and the {\it apparently standard regime}  at  some critical luminosities,
which are typically several tens of percents of  $L_{\rm E}$
(Kubota et al. 2001; Kubota \& Makishima 2004; Abe et al. 2005).
Around  this critical luminosity, yet another spectral state,
called {\it very high} state (VHS; \cite{Miyamoto1991}),
sometimes appears.
In this VHS, the spectrum becomes dominated by the power-law tail
presumably due to Comptonization,
even though the optically-thick disk emission
underlies the spectrum  Kubota et al. (2001; 2004).
\citet{kubota02} argued
that the power-law dominant state of ULXs is  analogous to the VHS of BHBs,
rather than to their classical low state.
This idea 
was also applied successfully to 
other power-law type ULXs (e.g., \cite{stobbart06}).
Then, the ULX transitions between the disk dominant and 
power-law (plus soft excess) type states
can be regarded as analogous  to the BHB transitions
between the {\it apparently standard} state and the VHS.
This solves the third problem,
because the transition luminosity, 
$\sim10^{40}~{\rm erg~s^{-1}}$, is now considered close to $L_{\rm E}$,
rather than to several percent of $L_{\rm E}$.

Finally, let us touch on the black hole mass in M81 X-9. 
If the observed spectra are securely confirmed as emerging from
the standard accretion disk, we can estimate the true inner radius,
$R_{\rm in}$,  from the measured, $r_{\rm in}$, through the corrections employed in
\S 3.1. We can then identify $R_{\rm in}$  with the last stable orbit,
to get an estimate on the black hole mass. However, now that the
accretion disk in M81 X-9 has been confirmed to be significantly
deviated from the standard disk, the correction from the apparent
radius, $r_{\rm in}$,  to the true physical radius, $R_{\rm in}$, would become highly
model dependent. Even if we somehow derived $R_{\rm in}$, we may not be
justified in identifying it with $6R_{\rm g}$ in an extreme slim disk
condition. We thus do not attempt to estimate the black hole mass
based on the radius parameter obtained with the $p$-free fit.

Instead of using the radius parameter, a much safer argument
on the black-hole mass can be made based on the luminosity.
To keep the black hole mass to a normal stellar black hole of $\sim10$--$20~M_\odot$,
the observed luminosity is a factor 7--10 higher than its Eddington limit.
Of course, such a highly super-Eddington luminosity could be possible from a theoretical viewpoint (e.g., Ohsuga et al. 2005). 
However, it would be more natural to consider that the object is radiating at close to its Eddington luminosity, considering that highly super-Eddington luminosities have not been observed from Galactic/Magellanic BHBs even if the disk deviate from the standard disk supposedly.
%
In this case, the black hole mass is estimated as $\sim100~M_\odot$.
This exceeds a typical upper end of black hole mass 
which can reasonably form from models of single stellar collapse, 
and may suggest some kinds of merging (Ebisuzaki et al. 2001; Belczynski et al. 2004).

\begin{figure}
  \begin{center}
    \FigureFile(80mm,80mm){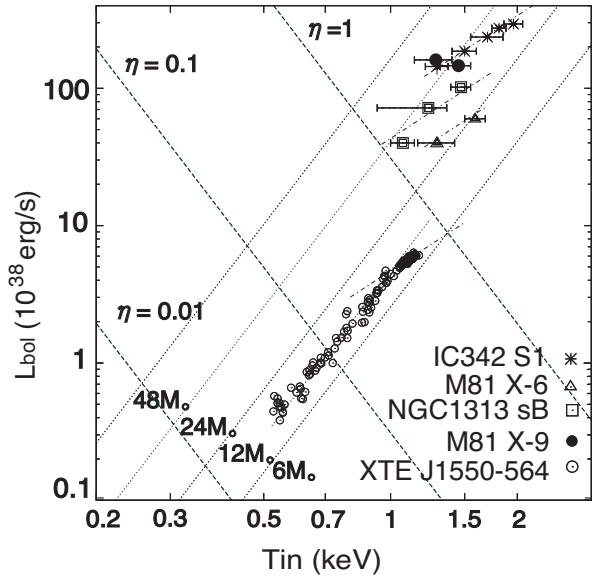}
  \end{center}
  \caption{The calculated $L_{\rm bol}$ plotted against $T_{\rm in}$, both
obtained from the MCD + power-law model.
The dotted and dot-dashed lines represent 
the relations of $L_{\rm bol} \propto T_{\rm in}^4$
and $L_{\rm bol} \propto T_{\rm in}^2$, respectively.
The dashed lines are the relation of 
$L_{\rm bol} \propto T_{\rm in}^{-4}$ (constant Eddington ratio $\eta\equiv L_{\rm bol}/L_{\rm Edd}$).
All points of M81 X-9 refer to the present paper.
Those of other ULXs, IC342 source 1, M81 X-6 and NGC 1313 Source B,
are obtained with ASCA (\cite{Mizuno2001}; \cite{Makishima2000}).
The data of the Galactic BHB, XTE~J$1550-564$, are obtained with {\it RXTE} (\cite{Kubota2004}). 
The disk inclinations of the ULXs are assumed as $60^{\circ}$.}
  \label{fig:Tin-Lbol}
\end{figure}

\bigskip

\section*{Acknowledgement}
We are grateful to the XMM-Newton science operation center 
and the center for planning and information systems at ISAS/JAXA for 
making the data publicly available.
We would like to thank Y. Maeda and K. Iwasawa for their advices on analysis of off-axis sources.
We also thank T. Tanaka and R. Miyawaki for helpful discussions.
We are grateful to T. Mizuno for useful comments on this paper.
A.K. is supported by a special 
postdoctoral researchers program in RIKEN.
The present work is supported in part by Grant-in-Aid
for Priority Research Areas,
No.14079201 and No.1703011, 
from Ministry of Education, Culture, Sports, Science and Technology of Japan.
It is also supported in part by the RIKEN's budget on
"Investigations of Spontaneously Evolving Systems".

\appendix
\section*{The analysis of SNR G21.5-09}

\begin{figure}[htb]
  \begin{center}
    \FigureFile(80mm,80mm){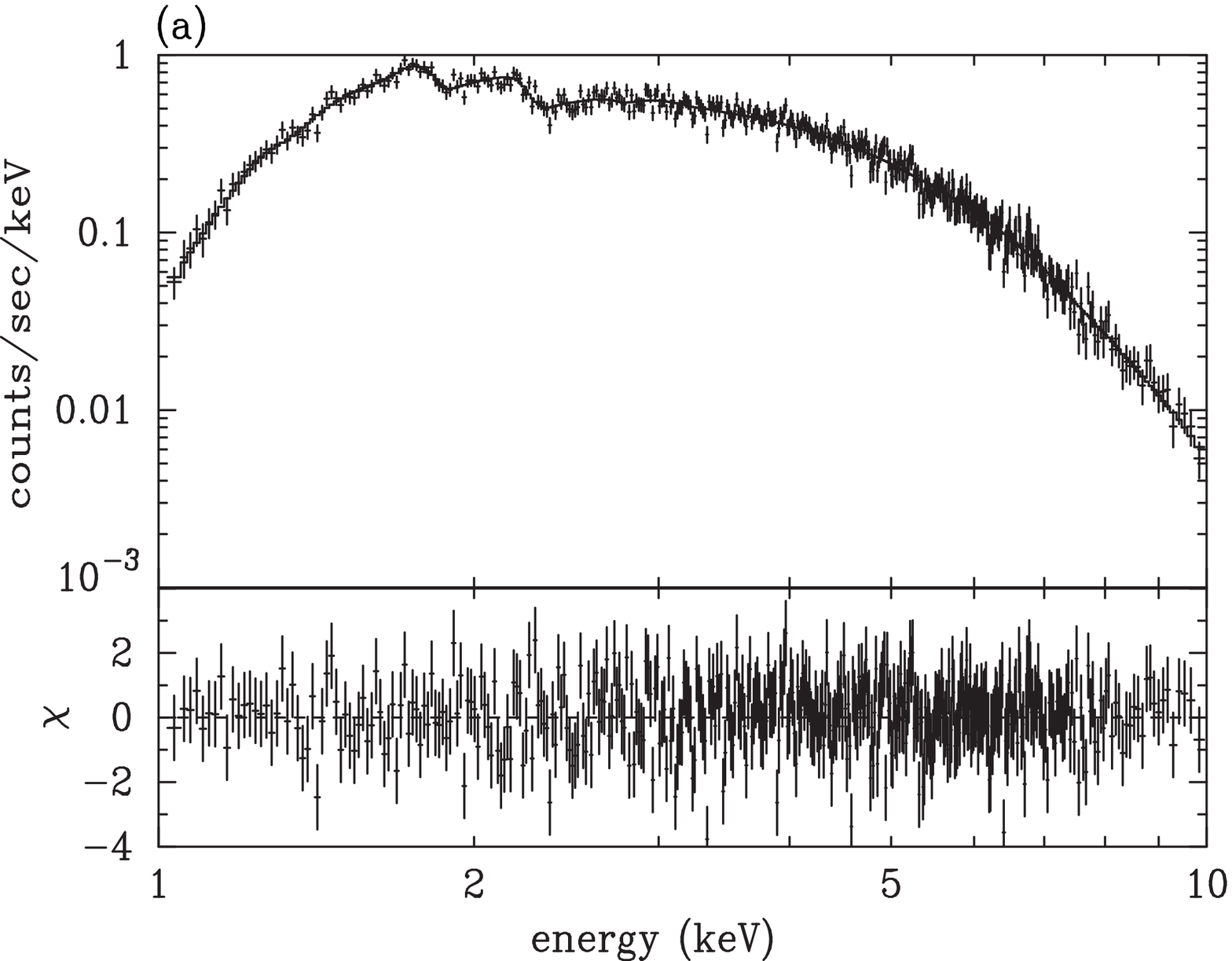}

    \FigureFile(80mm,80mm){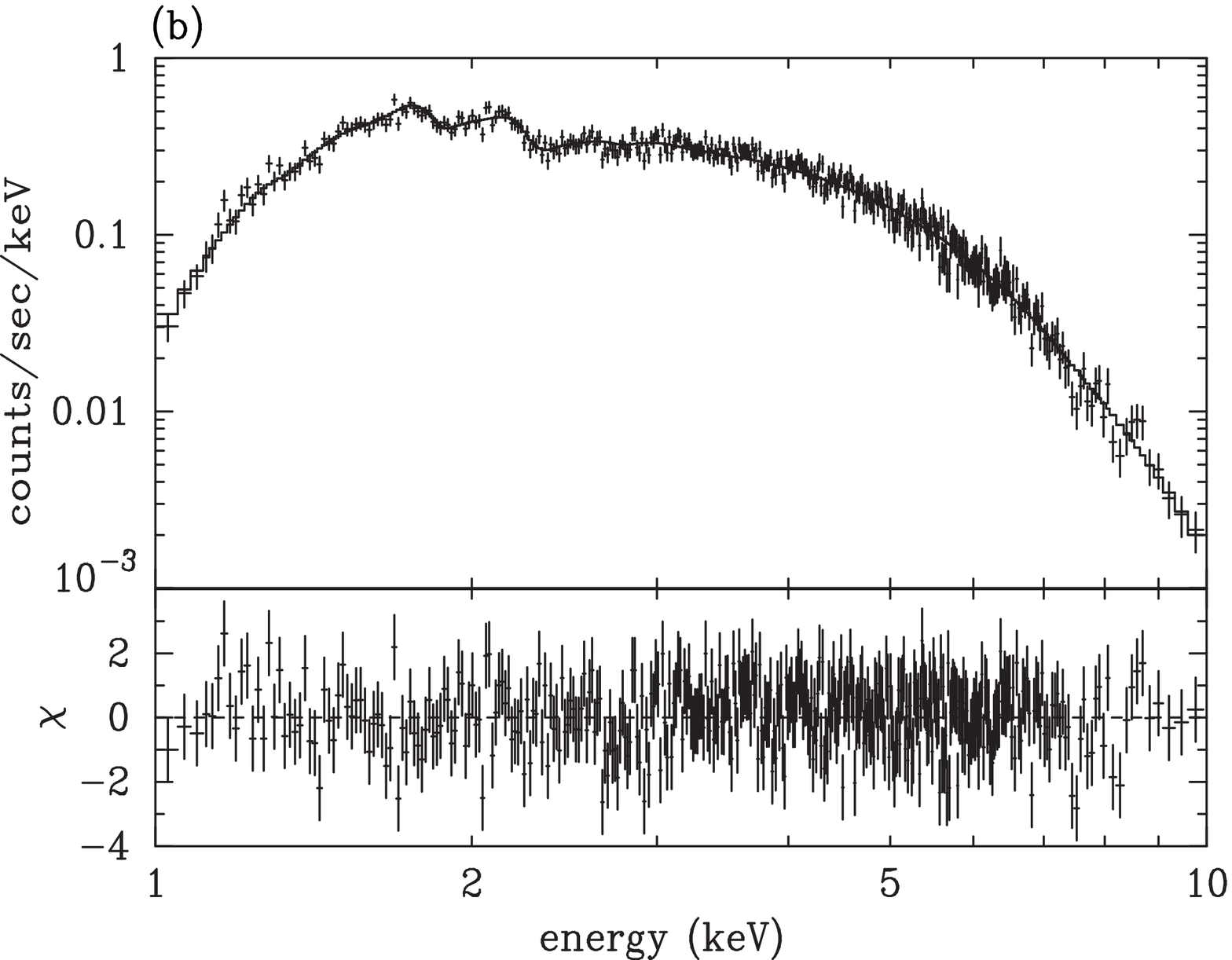}
    \FigureFile(80mm,80mm){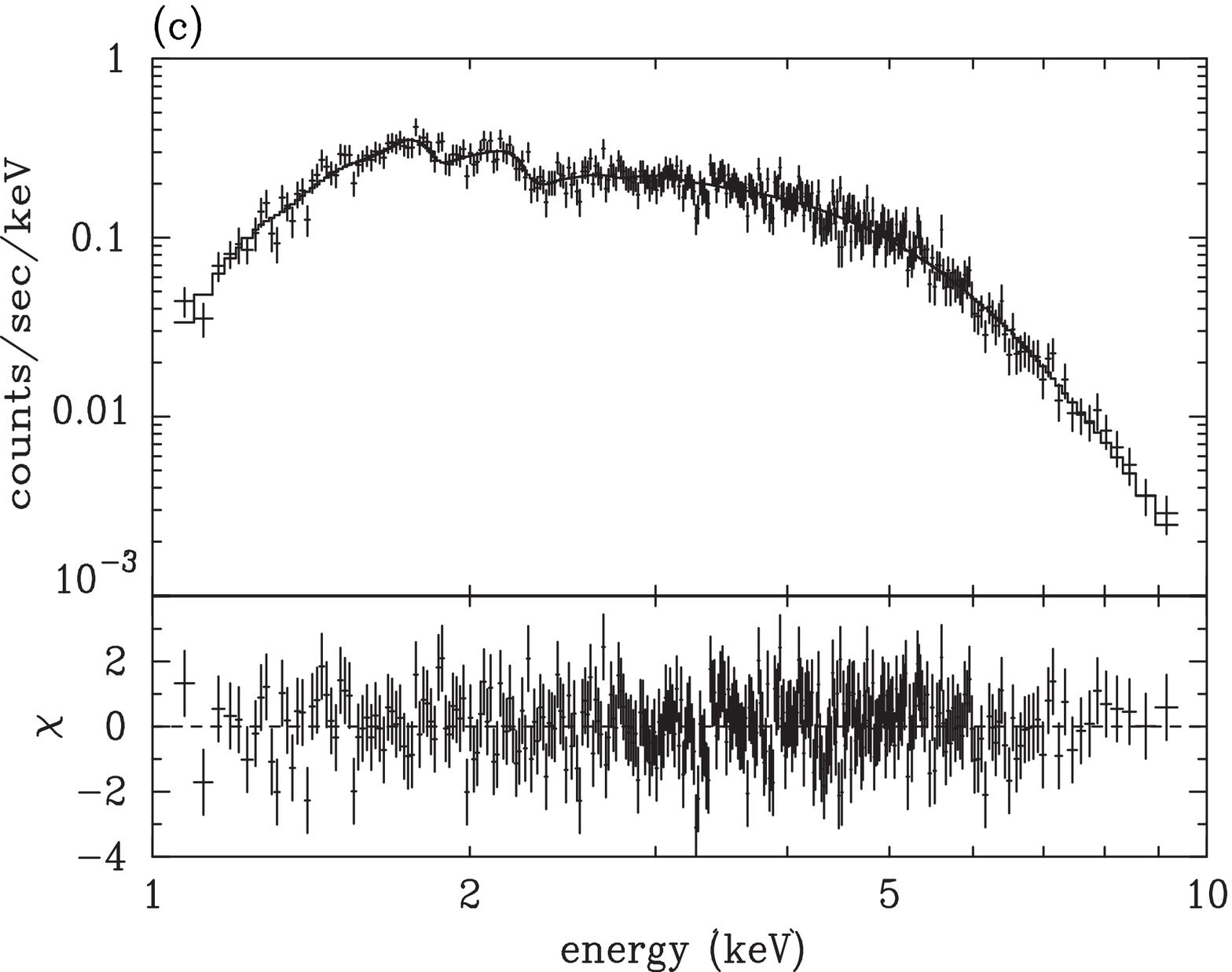}
  \end{center}
  \caption{MOS-1 spectra of G$21.5-09$ of acquired in; (a) an on-axis pointing, and off-axis pointings with 
  offsets of (b) $11^\prime\!.12$ and (c) $13^\prime\!.02$.
  The best fit power-law model is superposed on the data.}
  \label{fig:G21.5-09_spec}
\end{figure}

\begin{figure}
  \begin{center}
    \FigureFile(80mm,80mm){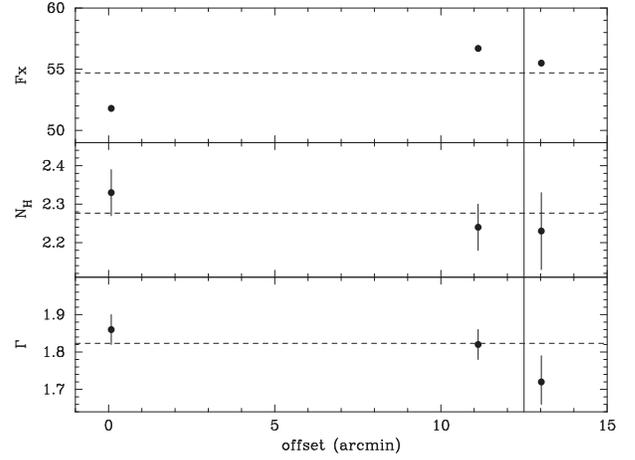}
  \end{center}
  \caption{The best-fit power-law model parameters of G21.5-09, plotted
 against the off-axis angle. From top to bottom, plotted are the 1--10
 keV flux (in $10^{-11} \rm{erg~s^{-1}~cm^{-2}}$), the hydrogen column
 density for the photoelectric absorption (in $10^{22} \rm{cm^{-2}}$),
 and the photon index. 
The line drawn at \timeform{12'.5} indicates the off-axis distance of
 M81 X-9 owing the 2001 April 22 observation. }
  \label{fig:G3para}
\end{figure}

In order to examine the reliability of spectral and photometric results
on highly off-axis XMM-Newton sources, we analyzed archival 
XMM-Newton data of the supernova remnant G21.5-09, acquired in off-axis and on-axis pointings.
This source, one of Crab-like supernova remnants, has been observed many
times with XMM-Newton for calibrating the mirror vignetting (\cite{Lumb2004}).
The observations were performed on 2000 April 7 for 17 ks, 2000 April 11
for 25 ks, and 2001 April 1 for 15 ks, with off-axis angle of
\timeform{0'.08}, \timeform{11'.12} and \timeform{13'.02}, respectively.
We analyzed these data sets in the same way as for M81 X-9. 
Figure \ref{fig:G21.5-09_spec} and Table \ref{ta:G21.5-0.9results} show
the 1--10 keV spectra, and the parameters obtained from single power-law
fits to them. 
Our results on the on-axis data are consistent with those reported by \citet{Warwick2001}.
As shown in Figure \ref{fig:G3para}, the off-axis spectral parameters
are consistent with the on-axis ones and the absorbed 1--10 keV flux is
reproduced to within 10\%.

\begin{table*}[htb]
  \caption{Spectral fitting parameters for G21.5-0.9 (1--10 keV).}
  \label{ta:G21.5-0.9results}
  \begin{center}
  \begin{tabular}{lcccc} 
\hline
   offset & ${N_{\rm{H}}}$\footnotemark[$*$] 
   & $\rm{\Gamma}$\footnotemark[$\dagger$] 
   & $F_{\rm{X}}$\footnotemark[$\ddagger$] & $\rm{{\chi}^2 / d.o.f.}$ \\ 
\hline\hline
  \timeform{0'.08} & 23.3 $\pm0.6$  & 1.86 $\pm0.04$ & 51.8 
           & 446.5/434 \\ 
  \timeform{11'.12} & 22.4 $\pm0.6$  & 1.82 $\pm0.04$  & 56.7 
           & 422.6/399 \\  
  \timeform{13'.02} & 22.3 $\pm1.0$  & 1.72 $_{-0.06}^{+0.07}$  
           & 55.5 & 298.8/323 \\  
\hline
   \multicolumn{5}{l}{\footnotesize
	\par\noindent
	\footnotemark[$*$] Hydrogen column density for the photoelectric 
            absorption, in units of $\rm{10^{21}~cm^{-2}}$. }\\
   \multicolumn{5}{l}{\footnotesize
	\par\noindent
	\footnotemark[$\dagger$] Photon Index~$\Gamma$. }\\
   \multicolumn{5}{l}{\footnotesize
	\par\noindent
	\footnotemark[$\ddagger$] The uncorrected absorption flux in the 
            range of 1-10 keV, in units of 
            $\rm{10^{-12}~erg~s^{-1} cm^{-2}}$. }\\
    \end{tabular}
  \end{center}
\end{table*}

\end{document}